# Research Contribution of major Centrally Funded Institution Systems of India

**Anurag Kanaujia, Prashasti Singh, Abhirup Nandy, Vivek Kumar Singh[1]**

Department of Computer Science, Banaras Hindu University, Varanasi-221005, India.

**Abstract**

India is now among the major knowledge producers of the world, ranking among the top 5 countries in total research output, as per some recent reports. The institutional setup for Research & Development (R&D) in India comprises a diverse set of Institutions, including Universities, government departments, research laboratories, and private sector institutions etc. It may be noted that more than 45% share of India's Gross Expenditure on Research and Development (GERD) comes from the central government. In this context, this article attempts to explore the quantum of research contribution of centrally funded institutions and institution systems of India. The volume, proportionate share and growth patterns of research publications from the major centrally funded institutions, organised in 16 groups, is analysed. These institutions taken together account for 67.54% of Indian research output during 2001 to 2020. The research output of the centrally funded institutions in India has increased steadily since 2001 with a good value for CAGR. The paper presents noteworthy insights about scientific research production of India that may be useful to policymakers, researchers and science practitioners in India. It presents a case for increased activity by the state governments and private sector to further the cause of sustainable and inclusive research and development in the country.

**Keywords**: Indian Research, Indian Science, Knowledge Production, Scientific Research.

## Introduction

The Indian research and development (R&D) system is constituted of various organisations, such as universities, government research laboratories, autonomous organizations, private research laboratories and centres etc. Recent information[2] from University Grants Commission (UGC) suggests that there are 1043 universities, and about 40,000 affiliated colleges in the country. Currently, this system consists of 54 Central, 429 State, 125 Deemed, and 380 Private Universities and more than 150 Institutes of National importance. These cater to activities related to a wide variety of disciplines (such as arts, languages, sciences, social sciences, humanities etc.) and receive financial support from different sources. In addition to these, there are well established institutional systems which have significant contributions in the national R&D output. These include the laboratories/centres under Council for Scientific and Industrial Research (CSIR), Defence Research and Development Organization (DRDO), Indian Council for Agricultural Research (ICAR), Department of Atomic Energy (DAE) etc. There are various sources of R&D funding, as well as various R&D models which are followed by these organisations. The major portion of the funding for R&D, however, is provided by the central and state governments. Their contributions account for a total of 51.8 percent of total annual gross expenditure of R&D **(Figure 1)** of the country, while Private sector accounts for approx. 37 percent of GERD. In this context, the proportionate contribution of the centrally funded and private funded institutions and institution systems in the total research output of India is not well-known.

---

[1] Corresponding author. Email: vivekks12@gmail.com

[2] as on 18.06.2021, UGC (https://www.ugc.ac.in/oldpdf/consolidated%20list%20of%20all%20universities.pdf)

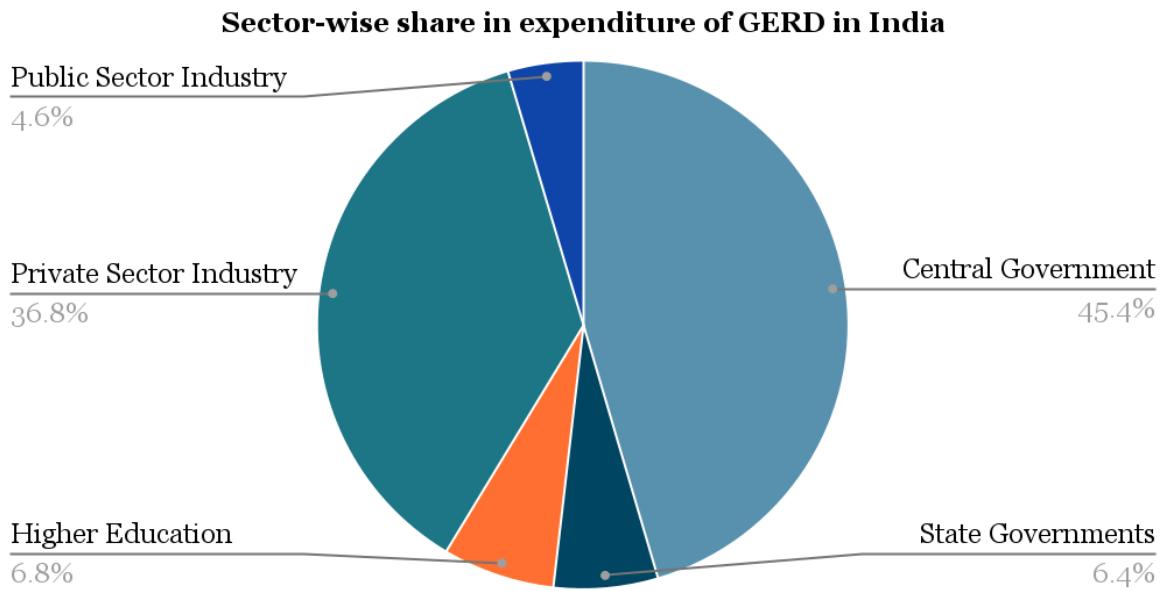

**Figure 1.** Relative percentages of GERD shares of the different sectors STI in India, 2018-19
Source: NSTMIS, Department of Science & Technology, GoI (2021)

Several previous studies have explored the research productivity of the various institutional groups such as Central Universities (Marisha, Banshal and Singh 2017, Basu *et al.* 2016), Indian Institute of Technology (IITs) (Banshal *et al.* 2017, Prathap and Gupta 2009, Prathap 2013), National Institute of Technology (NITs) (Banshal, Solanki and Singh 2018, Bala and Kumari 2013), Indian Institute of Science Education and Research (IISERs) (Solanki, Banshal and Singh 2016), Private universities (Banshal, Singh and Mayer 2019, Prathap and Sriram 2017), and research intensive higher education institutions (Prathap 2014, Nishy *et al.* 2012). Some other studies also focused on analysing research performance of India at an overall or broader level in different contexts such as, (Rajan, Swaminathan and Vaidhyasubramaniam 2018, Prathap 2018, Singh *et al.* 2022). To the best of our knowledge, there are no studies measuring and comparing the research output of all of the centrally funded institutions to the overall research output from India. Therefore, this study attempts to present an analytical account of the research output from the major centrally funded institutions and institution systems of India.

**Overview of the Centrally funded Institution Systems**

The Indian centrally funded institutional system comprises of a diverse set of institutions and institution systems. For the purpose of this analysis, we have categorised them into three top level categories: (i) Ministries, Departments and Autonomous Organizations under them, (ii) Higher Education Institutions funded by central government, and (iii) Councils and Agencies maintaining different institutions (**Figure 2**). Although CSIR and DRDO are also accorded department status, we have put them under third category as they maintain a large number of institutions and laboratories. Among these, four ministries/ departments have a total of 71 institutions (Ministry of Earth Sciences (9), Department of Atomic Energy (17), Department of Science and Technology (23), and Department of Biotechnology (22)). The second category includes Higher Education Institutions and comprises of 138 entities (Central Universities (36), Indian Institute of Technology (23), National Institute of Technologies (31), Indian Institute of Management (20), Indian Institute of Science Education and Research (6), All India Institute of Medical Sciences (8) and National Institute of Pharmaceutical Education and Research (7)). The third category includes five councils/agencies have 439 institutions (Council of Scientific and Industrial Research (43), Indian Council of Agricultural Research (95), Institutes of Indian Council of Medical Research (31), Indian Space Research Organisation (16) and Defence Research and Development Organisation (61)). It may be noted that some categories do not exhaustively include all institutions in that category as institutions that do not have significant amount of research output included in the Web of Science database are left out. For example, under Central University category, Universities like Maulana Azad National Urdu University (MANUU), Mahatma Gandhi Antarrashtriya Hindi Vishwavidalaya (MGAHV) etc. are not included in the Web of Science database due to low volume of scientific research output. The detailed list of institutions included in each group is given in **Appendix 1**.

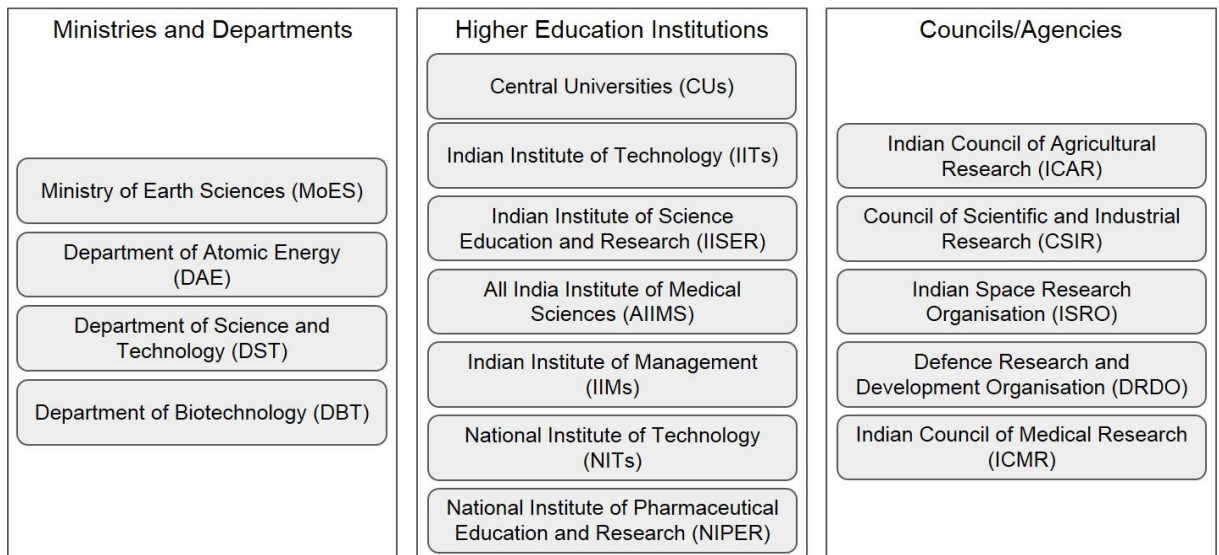

**Figure 2.** Indicative categorisation of groups of institutions looked at in the study.

**Objectives**

The present study attempts to answer the following research questions:

**RQ1:** What is the overall contribution of centrally funded institutions and institution systems to the total research output of India during 2001 to 2020?

**RQ2:** Which institutions and institution systems show a higher rate of growth of research output, measured in terms of volume and proportionate share?

**Data & Method**

The data for analysis in this paper corresponds to research publication data (article and review document types) for various centrally funded institutions and institution systems of India. The Web of Science (WoS) database has been used as the source of research publication data. For the purpose of obtaining data, a list of centrally funded institution systems has been prepared after checking the institutional websites and other relevant reports and documents. Data for a total of 16 institution systems/groups consisting of 454 individual institutions is obtained. It may be noted that Web of Science does not have an appropriate institution system grouping for majority of the institutions. Therefore, the data for the institutions was downloaded individually. The data for the period 2001 to 2020 was downloaded for all the individual institutions as well as for India as a whole. Data download was performed in the month of April 2022. The following search queries were used for this purpose: (a) PY=(2001-2020) AND CU="India" AND DT=("Article" OR "Review") to collect research output data for India, and (b) PY=(2001-2020) AND CU="India" AND OG=X AND DT=("Article" OR "Review") to collect data on the research output of individual institutions. Here where, PY stands for Publication Year, OG stands for Organization, DT stands for Document Type and CU stands for Country, which is India in our case. X is the name of the institution queried for.

The publication data obtained from WoS was analysed by writing programs in Python. The standard indicators of total research output, proportionate share, and Compounded Annual Growth Rate (CAGR) was computed for all the institution groups. The Compounded Annual Growth Rate (CAGR) of institutions is calculated as follows:

$$\text{CAGR} = \left(\left(\frac{V_{final}}{V_{begin}}\right)^{\frac{1}{t}} - 1\right) * 100$$

where, $V_{final}$ is the number of publication records in the year 2019, $V_{begin}$ is the number of publication records in the year when the first research output of the institution is seen, and t is the time period between the first and latest publications in years. The various results were computed for the 16 major identified institution systems. The proportionate contribution of each institution system to India's total research output was identified. The data was divided into four different blocks of 5-year each. The patterns of growth and the CAGR value for all institution systems were computed and plotted. Next, the change in proportionate contribution of each institution system during the four blocks is observed.

**Research Contribution of Centrally funded institution systems**

The analytical results described below present the research output volume of each institution system along with its proportionate share to India's research output. Thereafter, the patterns of growth of research output of the institution systems are presented.

*Research Output volume of each institution system between 2001 and 2020*

The total number of publications, percentage share to the national output and Compound Annual Growth Rate (CAGR) of all the institution systems considered is presented in **Table 1**. The IIT system is the largest contributor with a total of 1,52,276 papers during 2001 to 2020. It is followed by CSIR (99,430), Central Universities (97,524), DAE (77,819), NIT (46,034) and ICAR (44,733). The 23 institutions of DST combined together contribute 33,818 papers. In terms of percentage share to India's total output during the period, IIT system contributes 15.8%, followed by CSIR with 10.32% and CU with 10.12%. The IISER system and NIT system record very impressive value of CAGR. It may be relevant here that institution systems like CSIR, ISRO and DRDO engage in technology development activities in addition to research activities. Similarly, the higher education institution systems like IIT, CU, NIT etc engage in teaching activities in addition to research. The relative emphasis of different institution systems on research and other activities is different. Further, the institution systems vary significantly in terms of number of full-time researchers and availability of resources. Therefore, the results should not be seen as an effort to compare the research output of these institutions with each other rather they should be seen in terms of their contribution to India's total research output for the period.

**Table 1**: Research Output of Institution Systems during 2001 to 2020.

| Institution System | Number of Institutions | TP (2001-2020) | Percentage share to National Output (%) | CAGR (%) |
|---|---|---|---|---|
| IIT | 23 | 152,276 | 15.8 | 11.56 |
| CSIR | 43 | 99,430 | 10.32 | 6.38 |
| CU | 36 | 97,524 | 10.12 | 9.82 |
| DAE | 17 | 77,819 | 8.07 | 7.22 |
| NIT | 31 | 46,034 | 4.78 | 23.05 |
| ICAR | 95 | 44,733 | 4.64 | 8.67 |
| DST | 23 | 33,818 | 3.51 | 5.61 |
| ISRO | 19 | 22,666 | 2.35 | 6.51 |
| AIIMS | 8 | 15,654 | 1.62 | 9.00 |
| DRDO | 61 | 13,285 | 1.38 | 6.12 |
| DBT | 24 | 13,262 | 1.38 | 8.74 |
| IISER | 6 | 11,556 | 1.2 | *57.80 |
| ICMR | 31 | 11,061 | 1.15 | 7.79 |
| MoES | 10 | 5,786 | 0.6 | 12.18 |
| IIM | 20 | 3,248 | 0.34 | 15.26 |
| NIPER | 7 | 2,732 | 0.28 | 10.91 |

**Note**: Percentage share is corresponding to Indian Research Output of 963,709 publications (Article + Review) during 2001-2020. *CAGR taken from 2006 to 2020 as 1st instance of publications made in 2006.

The percentage share of each institution system in the overall national output is calculated and visualised on a pie chart for a better understanding **(Figure 3)**. The top six systems contribute greater than 50 percent of the total output. All the centrally funded Institution systems combined produce 67.54% of the total research output of India during the period. Among the top contributors, IITs, CSIR laboratories, and Central Universities have greater than 10 percent shares each. It is also interesting to observe that NIT, IISER and IIM systems, although contributing only 4.78 percent, 1.2 percent and 0.34 percent, have a very steep rise in their contributions which is reflected in their high CAGRs (23.05, 57.80 and 15.26 percent).

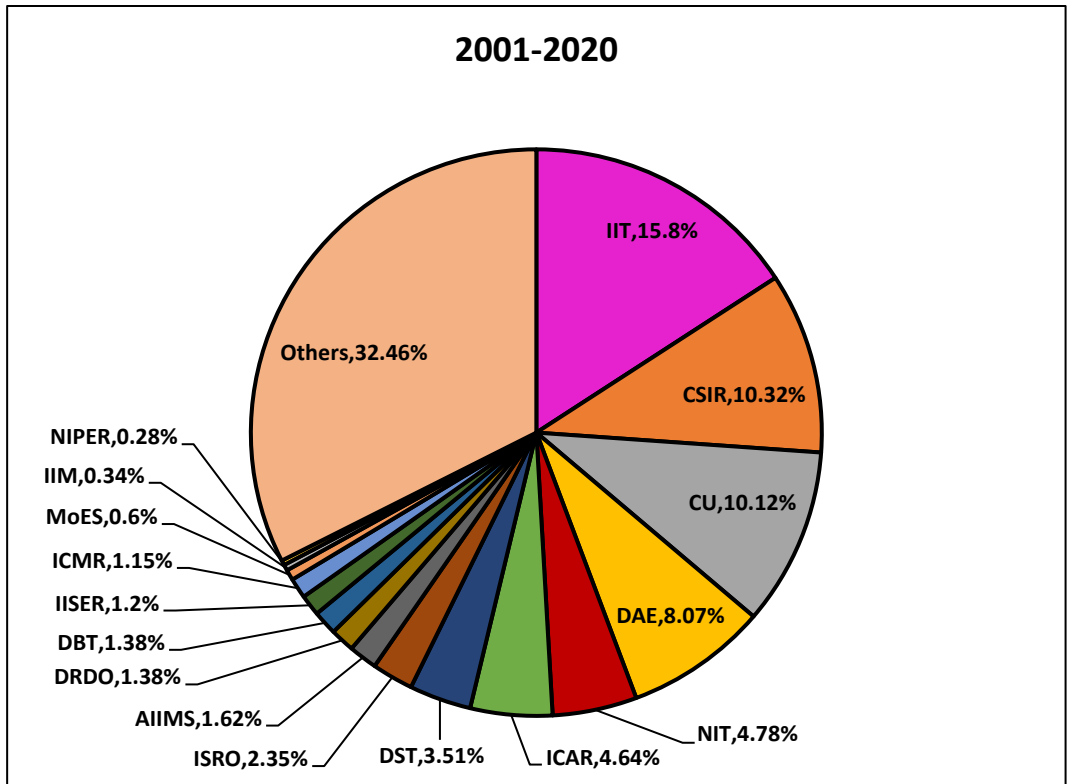

**Figure 3.** Percentage contributions of selected institution systems in India's Research output from 2001 to 2020. The highest contributions are made by IITs, CSIR Laboratories, Central Universities, all contributing more than 10 percent.

*Patterns of Growth in Research output*

To understand the patterns of growth in research output from each institution system, a set of bar charts are shown for four blocks of 5-year each for all the institution systems (**Figure 4**). A look at individual research output of the group of institutions shows growing research output in all the cases. IITs, NITs, IISERs, and IIMs have rapid growth in their research output, with a two-fold increase in the number of publications between 2011-15 and 2016-2020. While the publications of other institutions have increased, NIPERs have stagnated in the period between 2011-2020. IIT, CSIR, CU and DAE systems shows consistent growth in all the four blocks. ICAR, DST, ISRO and DBT systems also show overall growth in all the four blocks. Thus, in overall terms, majority of the institution systems record noticeable growth in research output during the 2001-20 period.

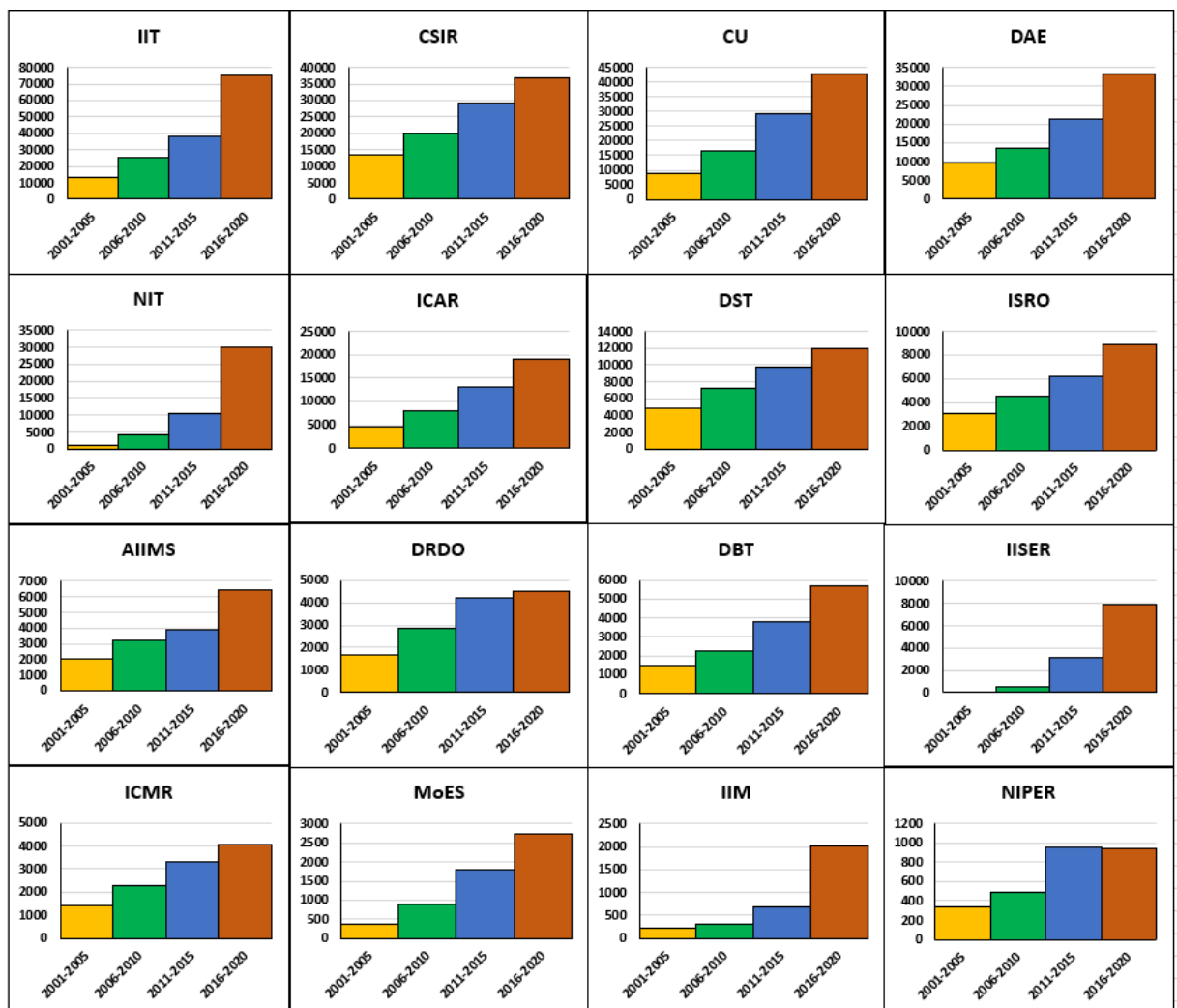

**Figure 4.** Research output of each group of institutions during the period of 2001-2020.

To further study the trends in the research output and percentage contribution of the different institution systems, the proportionate share of research contribution of each institution system to India's total output is shown in **Figure 5**. The pie charts are shown for four different time period blocks, 2001-05, 2006-10, 2011-15, and 2016-20. It can be observed that the proportionate share of IIT systems has increased from 12.56% in 2001-05 to 18.72% in 2016-20. On the other hand, the proportionate contribution of CSIR system has decreased continuously during the period, decreasing to 9.14% in 2016-20 from 12.43% in 2001-05. The CU system has also shown a marginal increase in proportionate share during first three blocks, and a stagnation at the same level in fourth block. DAE system showed a mixed pattern of decline and then a growth in proportionate share. NIT system has improved its percentage share significantly, increasing of a mere 1.03% in 2001-05 to 7.5% in 2016-20. This is a very impressive growth of research output for a system. ICAR system's proportionate share remained constant at the same level during the period. The proportionate share has remained largely unchanged in case of other institution systems too namely, DAE, ICAR, ISRO, AIIMS, DRDO, DBT, ICMR and NIPER. Over the twenty-year period, the contribution of the rest of the institutions has reduced from 37.54% in 2001-2006 to 27.3% in 2016-2020, implying that the central institution systems combined have increased their contribution.

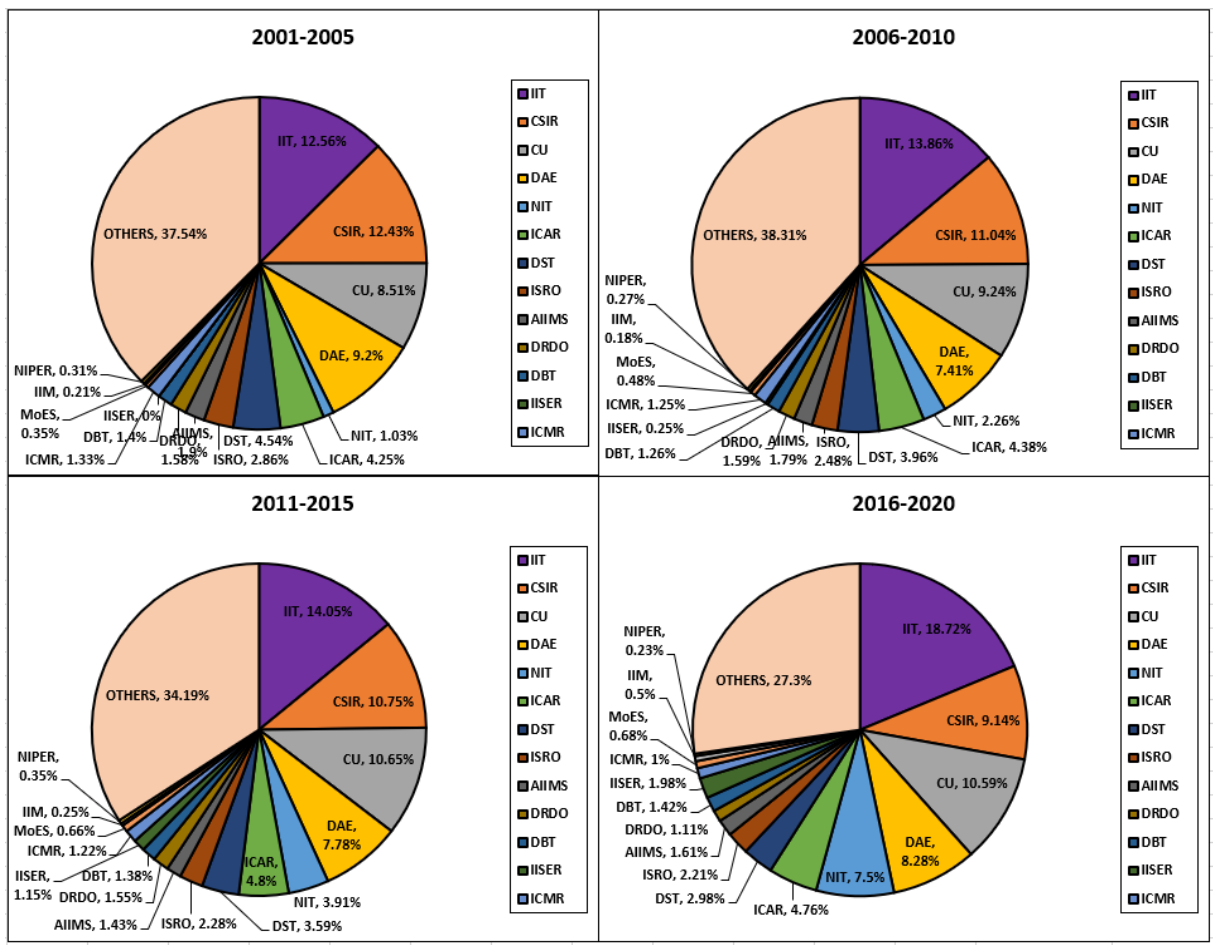

**Figure 5.** Proportional research publications from centrally funded groups of institutions in five-year intervals starting from 2001. Centrally funded institutions contribute to more than sixty percent of the research publications.

**Discussion**

The article analysed the research contribution of major centrally funded institution systems to India's total research output during 2001 to 2020. It is observed that research contribution of centrally funded institutions has increased from 62.46% in 2001-2006 to 72.7% in 2016-2020. Almost all major institutions have recorded growth in their research output volume during the period. However, in terms of proportionate share, some institutions (such as IITs and NITs) have shown significant growth, while the proportionate share of some other institutions (such as CSIR and DRDO) has decreased a bit. There could be two probable explanations for this. First, since more institutions have been added to some of the systems like IIT and NIT in the period, their total research output has increased at a much faster rate. Second, most of the institution systems have also recorded a genuine growth in research productivity in this period. Thus, in overall terms the research output volume and proportionate share of the major centrally funded institution systems have increased.

The relatively low proportionate share of institution systems like DRDO, ISRO, DST, DBT, ICMR etc. is an interesting point to observe. Unlike what one may expect, these institutions are not producing that high amount of research output. This may be explained from the fact that these organizations engage in different kind of research and development (R&D) activities, all of which does not result into a research paper. In this sense the paper has a limitation that it only considers the research papers and other R&D outputs (such as patents, technologies developed etc.) are not factored. A more detailed analysis of the research contribution of the institutions would thus need data about patents and technologies developed. Interestingly, the institutions of higher learning (such as IITs, NITs, CUs, IISERs etc.) have produced more research papers. This indicates that the Indian higher education institutions are now more seriously engaging in research activities, which quite often lead to research papers as outputs. Irrespective of these observations, one may note that the major centrally funded institution systems of India contributed approximately 2/3$^{rd}$ of total research output of India. The combined output of all other institutions (under various state governments as well as private sector) is quite less (just about 1/3$^{rd}$). The number of the institutions under state government and private sector taken together will be many times more than the total number of centrally funded institutions. Yet their research output is quite less. Thus, the results indicate that these centrally funded institutions have an important role to play in India's research and development activities, and that state governments should strive to promote more research and development activities in their institutions.

In the recent times, the Government (particularly central government) has taken various initiatives to increase allocation in the field of Science and Technology such as successive increase in plan allocations for scientific departments, setting up of new institutions for science education and research, creation of Centres of Excellence and facilities in emerging and frontline areas of S&T in academic and national institutions, supporting Mega Facilities for Basic Research, launching of new fellowships, substantial grant to potential scientists through extramural research funding, scaled up funding in the new areas such as Clean Energy and Water including Energy Efficiency, Clean Coal Technology, Smart Grids, Methanol, Desalination, Genome Engineering Technology, climate change research, National Supercomputing Mission (NSM), National Mission on interdisciplinary Cyber Physical System (ICPS) etc., promotion of innovation, entrepreneurship and start-ups grant for young scientists, Funds for Improvement of S&T Infrastructure (FIST), encouraging public-private partnerships, fiscal incentives and support measures for enhancing the participation of industry in R&D, etc. These initiatives could have directly or indirectly helped in promoting research and development activities in Indian institutions and particularly the institutions under central government appear to be the major beneficiaries.

In light of this information, in India, the role of stakeholders such as government and funding agencies is important in ensuring higher productivity from centrally funded as well as increasing the contribution of state and private funded institutions in research. Institutions under state government constitute a large share and an improvement in research culture in those institutions will result in manifold increase of India's research output. It is equally important that government and private stakeholders work together in a collaborative and complementary manner so as to ensure holistic growth in R&D capabilities of Indian Institutions.

**Appendix**

The list of over 400 institutions categorised in 16 groups is given in Appendix 1.

# Appendix 1

## List of Institutions in Centrally Funded Institution Systems

| Name of Institution | TP (2001-2020) |
|---|---|
| **Indian Institute of Technology (IIT) System** | |
| IIT Kharagpur | 23246 |
| IIT Madras | 19984 |
| IIT Bombay | 19483 |
| IIT Delhi | 19226 |
| IIT Kanpur | 15452 |
| IIT Roorkee | 14150 |
| IIT Guwahati | 10499 |
| Indian School of Mines Dhanbad | 6509 |
| IIT BHU Varanasi | 6465 |
| IIT Hyderabad | 3218 |
| IIT Indore | 3174 |
| Bhubaneswar | 2191 |
| IIT Patna | 1849 |
| IIT Ropar | 1736 |
| IIT Gandhinagar | 1685 |
| IIT Mandi | 1536 |
| IIT Jodhpur | 991 |
| IIT Tirupati | 228 |
| IIT Palakkad | 196 |
| IIT Jammu | 142 |
| IIT Bhilai | 125 |
| IIT Goa | 105 |
| IIT Dharwad | 86 |
| **Council of Scientific and Industrial Research (CSIR) System** | |
| Indian Institute of Chemical Technology (IICT) | 10680 |
| Academy of Scientific & Innovative Research (AcSIR)* | 9973 |
| National Chemical Laboratory (NCL) | 8736 |
| National Physical Laboratory (NPL) | 5420 |
| Central Drug Research Institute (CDRI) | 4970 |
| National Institute Interdisciplinary Science & Technology (NIIST) | 3700 |
| Central Food Technological Research Institute (CFTRI) | 3495 |
| Central Electrochemical Research Institute (CECRI) | 3431 |
| Central Leather Research Institute (CLRI) | 3367 |
| Indian Institute of Chemical Biology (IICB) | 3111 |
| National Institute of Oceanography (NIO) | 3086 |
| Central Salt & Marine Chemical Research Institute (CSMCRI) | 2949 |
| National Geophysical Research Institute (NGRI) | 2504 |
| Centre for Cellular & Molecular Biology (CCMB) | 2477 |
| Central Glass & Ceramic Research Institute (CGCRI) | 2454 |
| Institute of Genomics & Integrative Biology (IGIB) | 2135 |
| Indian Institute of Toxicology Research (IITR) | 2104 |
| National Botanical Research Institute (NBRI) | 1979 |
| Institute of Minerals & Materials Technology (IMMT) | 1934 |
| National Metallurgical Laboratory (NML) | 1900 |
| Central Institute of Medicinal & Aromatic Plants (CIMAP) | 1862 |
| National Environmental Engineering Research Institute (NEERI) | 1804 |
| Indian Institute of Integrative Medicine (IIIM) | 1770 |
| Institute of Microbial Technology (IMTECH) | 1593 |
| North East Institute of Science & Technology (NEIST) | 1474 |
| Institute of Himalayan Bioresource Technology (IHBT) | 1421 |
| National Aerospace Laboratories (NAL) | 1414 |
| Indian Institute of Petroleum (IIP) | 1017 |
| Central Mechanical Engineering Research Institute (CMERI) | 974 |
| Central Scientific Instruments Organisation (CSIO) | 950 |
| Central Electronics Engineering Research Institute (CEERI) | 839 |
| Advanced Materials & Processes Research Institute (AMPRI) | 758 |
| Central Institute of Mining & Fuel Research (CIMFR) | 756 |
| Structural Engineering Research Center (SERC) | 618 |
| Central Building Research Institute | 422 |
| Madras Complex (CMC) | 352 |
| Fourth Paradigm Institute (CSIR 4PI) | 281 |
| Central Road Research Institute (CRRI) | 232 |
| National Institute of Science, Technology & Development Studies (NISTADS) | 163 |
| Human Resource Development Centre | 116 |
| National Institute of Science Communication & Information Resources (NISCAIR) | 114 |
| Open Source Drug Discovery | 50 |
| Unit for Research and Development of Information Products | 45 |
| **Central University System** | |
| University of Delhi | 18353 |
| Banaras Hindu University (BHU) | 16969 |
| Aligarh Muslim University | 9699 |
| University of Hyderabad | 7764 |
| Jawaharlal Nehru University | 6752 |
| Jamia Millia Islamia | 5408 |
| Visva Bharati | 4111 |
| Pondicherry University | 3920 |
| University of Allahabad | 3757 |
| Tezpur University | 3230 |
| North Eastern Hill University | 2259 |
| Dr. Harisingh Gour Vishwavidyalaya | 2009 |

| | | | | |
|---|---|---|---|---|
| Assam University | 1469 | | NIT Kurukshetra | 2064 |
| Hemwati Nandan Bahuguna Garhwal University | 1137 | | Dr Br Ambedkar NIT Jalandhar | 1889 |
| | | | NIT Silchar | 1729 |
| Babasaheb Bhimrao Ambedkar University | 1110 | | NIT Hamirpur | 1558 |
| Guru Ghasidas Vishwavidyalaya | 1076 | | NIT Raipur | 1381 |
| Central University of Rajasthan | 901 | | Maulana Azad NIT Bhopal | 1222 |
| Mizoram University | 864 | | NIT Agartala | 1098 |
| Manipur University | 825 | | NIT Patna | 777 |
| Central University of Punjab | 731 | | NIT Srinagar | 584 |
| Tripura University | 724 | | NIT Jamshedpur | 577 |
| Central University of Kerala | 556 | | NIT Meghalaya | 465 |
| Central University of Gujarat | 508 | | NIT Manipur | 353 |
| Central University of Tamil Nadu | 470 | | NIT Delhi | 324 |
| Central University of Jharkhand | 446 | | NIT Goa | 232 |
| Central University of Haryana | 323 | | NIT Arunachal Pradesh | 179 |
| Central University of Himachal Pradesh | 318 | | NIT Nagaland | 175 |
| Sikkim University | 307 | | NIT Uttarakhand | 174 |
| Rajiv Gandhi University | 305 | | NIT Puducherry | 161 |
| Central University of Jammu | 285 | | NIT Mizoram | 141 |
| Central University of South Bihar | 278 | | NIT Sikkim | 133 |
| Nagaland University | 229 | | NIT Andhra Pradesh | 109 |
| Indira Gandhi National Tribal University | 191 | | **Indian Council of Agricultural Research (ICAR) System** | |
| Central University of Karnataka | 142 | | | |
| Central University of Kashmir | 58 | | Indian Agricultural Research Institute | 7100 |
| The English and Foreign Languages University | 40 | | Indian Veterinary Research Institute | 4620 |
| | | | National Dairy Research Institute | 3049 |
| **Department of Atomic Energy System** | | | Central Marine Fisheries Research Institute | 1247 |
| Bhabha Atomic Research Centre | 21922 | | National Bureau of Plant Genetics Resources | 1124 |
| Tata Institute of Fundamental Research (TIFR) | 14558 | | | |
| | | | Central Institute of Fisheries Education | 1103 |
| Homi Bhabha National Institute | 7251 | | ICAR Research Complex for NEH Region | 1030 |
| Saha Institute of Nuclear Physics | 6294 | | Indian Institute of Horticultural Research | 941 |
| Indira Gandhi Centre for Atomic Research | 5786 | | National Rice Research Institute | 723 |
| Tata Memorial Centre | 3250 | | Central Institute of Freshwater Aquaculture | 713 |
| Institute of Physics, Bhubaneswar | 2956 | | Central Arid Zone Research Institute | 643 |
| Raja Ramanna Centre for Advanced Technology | 2910 | | Indian Grassland & Fodder Research Institute | 631 |
| National Institute of Science Education and Research | 2813 | | Central Sheep & Wool Research Institute | 627 |
| | | | National Bureau of Fish Genetic Resources | 563 |
| Institute for Plasma Research | 2587 | | National Bureau of Animal Genetic Resources | 557 |
| Institute of Mathematical Sciences | 2239 | | | |
| Variable Energy Cyclotron Centre | 2227 | | Indian Institute of Soil Sciences | 544 |
| Harish-Chandra Research Institute | 1893 | | Central Institute for Research on Goats | 543 |
| TIFR Centre for Interdisciplinary Sciences (TCIS), Hyderabad | 535 | | Central Soil Salinity Research Institute | 541 |
| | | | Central Institute of Fisheries Technology | 500 |
| Atomic Minerals Directorate for Exploration and Research | 361 | | Indian Institute of Wheat & Barley Research | 492 |
| | | | Central Avian Research Institute | 487 |
| TIFR Centre for Applicable Mathematics (CAM), Bengaluru | 214 | | National Institute of Animal Nutrition & Physiology | 479 |
| Tata Institute of Fundamental Research, Balloon Facility | 23 | | Indian Institute of Pulses Research | 477 |
| | | | Central Institute Brackishwater Aquaculture | 473 |
| **National Institute of Technology (NIT) System** | | | Central Institute on Post Harvest Engineering & Technology | 472 |
| NIT Rourkela | 5563 | | | |
| NIT Tiruchirappalli | 5140 | | Indian Institute of Rice Research | 470 |
| NIT Karnataka | 3308 | | Central Inland Fisheries Research Institute | 462 |
| NIT Durgapur | 2714 | | Central Potato Research Institute | 453 |
| NIT Warangal | 2569 | | Central Research Institute of Dryland Agriculture | 444 |
| Visvesvaraya NIT Nagpur | 2361 | | | |
| Sardar Vallabhbhai NIT Surat | 2357 | | Central Island Agricultural Research Institute | 401 |
| Malviya NIT Jaipur | 2342 | | | |
| Motilal Nehru NIT | 2262 | | | |
| NIT Calicut | 2093 | | | |

| Institute | Count | Institute | Count |
|---|---|---|---|
| Indian Agricultural Statistics Research Institute | 401 | National Institute of Agricultural Economics & Policy Research | 131 |
| Central Tuber Crops Research Institute | 398 | Indian Institute of Natural Resins & Gums | 123 |
| Indian Institute of Soil & Water Conservation | 388 | National Research Centre for Banana | 121 |
| Sugarcane Breeding Institute | 366 | National Research Centre for Integrated Pest Management | 112 |
| ICAR Research Complex for Eastern Region | 357 | Directorate of Weed Research | 109 |
| Indian Institute of Sugarcane Research | 344 | National Academy of Agricultural Research & Management | 104 |
| Indian Institute of Vegetable Research | 343 | National Research Centre on Seed Spices | 98 |
| Central Institute of Agricultural Engineering | 340 | Indian Institute of Agricultural Biotechnology | 91 |
| Vivekananda Parvatiya Krishi Anusandhan Sansthan | 332 | Directorate of Mushroom Research | 90 |
| Directorate of Poultry Research | 324 | Indian Institute of Oil Palm Research | 87 |
| National Bureau of Soil Survey & Land Use Planning | 310 | National Institute of Biotic Stresses Management | 84 |
| National Bureau of Agricultural Insect Resources | 309 | Directorate on Onion & Garlic Research | 83 |
| National Research Centre on Equines | 306 | National Research Centre for Pomegranate | 81 |
| Central Plantation Crops Research Institute | 298 | Central Tobacco Research Institute | 68 |
| Central Institute for Research on Buffaloes | 288 | Directorate of Cashew Research | 67 |
| Indian Institute of Spices Research | 281 | Central Institute for Women in Agriculture | 60 |
| National Institute of Veterinary Epidemiology & Disease Informatics | 279 | Indian Institute of Seed Research | 59 |
| Central Institute of Cotton Research | 261 | Directorate of Floricultural Research | 55 |
| National Research Centre on Camel | 259 | National Research Centre for Litchi | 46 |
| Indian Institute of Water Management | 256 | National Research Centre on Orchids | 40 |
| Central Research Institute for Jute & Allied Fibres | 249 | **Department of Science and Technology (DST) System** | |
| National Bureau of Agriculturally Important Microorganisms | 241 | Indian Association for the Cultivation of Science (IACS) - Jadavpur | 7892 |
| National Research Centre for Grapes | 220 | Jawaharlal Nehru Center for Advanced Scientific Research (JNCASR) | 5099 |
| National Research Centre on Yak | 215 | SN Bose National Centre for Basic Science (SNBNCBS) | 3031 |
| Indian Institute of Maize Research | 212 | Bose Institute | 3017 |
| National Research Centre on Mithun | 208 | Sree Chitra Tirunal Institute for Medical Sciences Technology (SCTIMST) | 2081 |
| Central Coastal Agricultural Research Institute | 206 | Raman Research Institute (RRI) | 2061 |
| Indian Institute of Millets Research | 206 | Indian Institute of Astrophysics (IIA) | 1842 |
| Directorate of Groundnut Research | 204 | International Advanced Research Centre for Powder Metallurgy & New Materials (ARCI) | 1101 |
| Central Institute of Subtropical Horticulture | 199 | Wadia Institute of Himalayan Geology (WIHG) | 1075 |
| Indian Institute of Oilseeds Research | 198 | Indian Institute of Space Science & Technology | 1009 |
| Indian Institute of Farming Systems Research | 191 | Birbal Sahni Institute of Palaeobotany (BSIP) | 987 |
| Central Citrus Research Institute | 185 | Indian Institute of Geomagnetism (IIG) | 953 |
| Directorate of Coldwater Fisheries Research | 185 | Agharkar Research Institute (ARI) | 860 |
| Indian Institute of Soybean Research | 185 | Aryabhatta Research Institute of Observational Sciences (ARIES) | 837 |
| Central Institute of Temperate Horticulture | 178 | Institute of Advanced Study in Science & Technology (IASST) | 678 |
| Directorate of Medicinal & Aromatic Plants Research | 174 | Centre for Nano & Soft Matter Sciences (CeNS) | 620 |
| Central Institute of Research on Cotton Technology | 172 | Institute of Nano Science & Technology (INST) | 583 |
| National Institute of Abiotic Stress Management | 170 | National Academy of Sciences, India (NASI) | 36 |
| Central Institute for Research on Cattle | 159 | Technology Information, Forecasting & Assessment Council (TIFAC) | 27 |
| National Institute of High Security Animal Diseases | 150 | National Innovation Foundation (NIF) | 16 |
| National Research Centre on Meat | 147 | Vigyan Prasar | 5 |
| National Research Centre on Pig | 147 | | |
| Central Agroforestry Research Institute | 141 | | |
| Central Institute for Arid Horticulture | 133 | | |

| Organisation | Count | Organisation | Count |
|---|---|---|---|
| Indian Science Congress Association (ISCA) | 4 | Defence Research & Development Laboratory (DRDL) | 299 |
| National Accreditation Board for Testing & Calibration Laboratories | 4 | Defence Research Laboratory (DRL) | 268 |
| **Indian Space Research Organisation (ISRO) System** | | Defence Laboratory (DLJ) | 252 |
| Department of Space DOS Government of India | 10979 | DRDO BU Center Life Sciences | 232 |
| Physical Research Laboratory India | 3421 | Snow & Avalanche Study Establishment (SASE) | 231 |
| Vikram Sarabhai Space Center VSSC | 2072 | Naval Physical & Oceanographic Laboratory (NPOL) | 214 |
| Space Applications Centre (SAC) | 1559 | Center for Fire Explosive & Environment Safety (CFEES) | 211 |
| National Remote Sensing Centre (NRSC) | 1053 | Defence Institute of High Altitude Research (DIHAR) | 187 |
| Indian Institute of Space Science Technology IIST | 1009 | Armament Research & Development Establishment (ARDE) | 174 |
| U R Rao Satellite Centre (URSC) | 717 | Laser Science & Technology Centre (LASTEC) | 161 |
| Indian Institute of Remote Sensing IIRS | 697 | Advanced System Laboratory (ASL) | 141 |
| National Atmospheric Research Laboratory (NARL) | 638 | Instruments Research & Development Establishment (IRDE) | 129 |
| Liquid Propulsion Systems Centre (LPSC) | 141 | Defence Institute of Bio-Energy Research (DIBER) | 116 |
| Semi-Conductor Laboratory (SCL) | 128 | Gas Turbine Research Establishment (GTRE) | 113 |
| ISRO Telemetry, Tracking & Command Network (ISTRAC) | 72 | Research Center Imarat (RCI) | 113 |
| Laboratory Electro-Optics Systems (LEOS) | 56 | Terminal Ballistics Research Laboratory (TBRL) | 110 |
| North Eastern-Space Applications Centre (NE-SAC) | 46 | Center for Military Airworthiness & Certification (CEMILAC) | 97 |
| Satish Dhawan Space Centre (SDSC) SHAR | 41 | Microwave Tube Research & Development Center (MTRDC) | 91 |
| ISRO Inertial Systems Unit IISU | 15 | Combat Vehicles Research & Development Establishment (CVRDE) | 78 |
| ISRO Propulsion Complex (IPRC) | 13 | Defence Bioengineering & Electromedical Laboratory (DEBEL) | 77 |
| Master Control Facility (MCF) | 8 | Naval Science & Technological Laboratory (NSTL) | 74 |
| Development & Educational Communication Unit DECU | 1 | Aeronautical Development Establishment (ADE) | 70 |
| **All India Institute of Medical Sciences (AIIMS) System** | | Scientific Analysis Group (SAG) | 65 |
| AIIMS New Delhi | 14183 | Defence Electronics Application Laboratory (DEAL) | 54 |
| AIIMS Bhubaneswar | 334 | Center for Artificial Intelligence & Robotics (CAIR) | 46 |
| AIIMS Jodhpur | 329 | Electronics & Radar Development Establishment (LRDE) | 46 |
| AIIMS Rishikesh | 261 | Institute of Systems Studies & Analysis (ISSA) | 44 |
| AIIMS Bhopal | 253 | Research & Development Establishment Engineers (R&DE) | 34 |
| AIIMS Patna | 159 | Research & Innovation Centre (RIC) | 33 |
| AIIMS Raipur | 90 | Defence Electronics Research Laboratory (DLRL) | 32 |
| AIIMS Nagpur | 45 | Defence Terrain Research Laboratory (DTRL) | 32 |
| **Defence Research and Development Organisation (DRDO) System** | | Proof & Experimental Establishment (PXE) | 29 |
| Defence Metallurgical Research Laboratory (DMRL) | 1925 | Defence Avionics Research Establishment (DARE) | 20 |
| Defence Research & Development Establishment (DRDE) | 1310 | Centre for Air Borne Systems (CABS) | 19 |
| Institute of Nuclear Medicine & Allied Sciences (INMAS) | 1095 | Integrated Test Range (ITR) | 19 |
| Defence Institute of Advanced Technology (DIAT) | 1011 | Recruitment & Assessment Centre (RAC) | 17 |
| Solid State Physics Laboratory (SSPL) | 823 | Defence Institute of Psychological Research (DIPR) | 14 |
| Defence Institute of Physiology & Allied Sciences (DIPAS) | 617 | | |
| Defence Food Research Laboratory (DFRL) | 597 | | |
| Advanced Centre Research in High Energy Materials (ACRHEM) | 522 | | |
| Defence Materials & Stores Research & Development Establishment (DMSRDE) | 518 | | |
| High Energy Materials Research Laboratory (HEMRL) | 441 | | |
| Naval Materials Research Laboratory (NMRL) | 395 | | |

| Institution | Count |
|---|---|
| DRDO Integration Centre (DIC) | 13 |
| Advanced Centre Energetic Materials (ACEM) | 11 |
| Aerial Delivery Research & Development Establishment (ADRDE) | 11 |
| Vehicle Research & Development Establishment (VRDE) | 10 |
| Defence Scientific Information & Documentation Centre (DESIDOC) | 8 |
| Advanced Numerical Research & Analysis Group (ANURAG) | 7 |
| SF Complex (SFC) | 7 |
| Centre High Energy Systems & Sciences (CHESS) | 6 |
| Centre of Millimeter Wave Semiconductor Devices & Systems (CMSDS) | 6 |
| Institute of Technology Management (ITM) | 5 |
| Centre of Excellence in Cryptology (CoEC) | 3 |
| Centre Advanced Systems (CAS) | 1 |
| Joint Cypher Bureau (JCB) | 1 |
| **Department of Biotechnology (DBT) System** | |
| International Center for Genetic Engineering & Biotechnology (ICGEB) | 1874 |
| Tamil Nadu Veterinary & Animal Sciences University | 1864 |
| National Institute of Immunology - India (NII) | 1537 |
| National Centre for Cell Science, Pune (NCCS) | 1470 |
| National Institute of Plant Genome Research (NIPGR) | 917 |
| Rajiv Gandhi Centre for Biotechnology (RGCB) | 906 |
| Centre for DNA Fingerprinting & Diagnostics (CDFD) | 852 |
| Institute of Life Sciences India (ILS) | 605 |
| National Brain Research Centre (NBRC) | 596 |
| Translational Health Science & Technology Institute (THSTI) | 577 |
| Institute for Stem Cell Biology & Regenerative Medicine - inStem | 427 |
| National Agri Food Biotechnology Institute (NABI) | 389 |
| Institute of Bioresources & Sustainable Development (IBSD) | 272 |
| Regional Centre for Biotechnology | 253 |
| National Institute of Biomedical Genomics (NIBMG) | 185 |
| National Institute of Animal Biotechnology (NIAB) | 161 |
| DBT-ICT Centre for Energy Biosciences, Mumbai | 94 |
| DBT IOC Centre for Advanced Bioenergy Research, Faridabad | 84 |
| DBT-AIST International Laboratory for Advanced Biomedicine (DAILAB) | 59 |
| DBT-Pan IIT Centre for Bioenergy | 44 |
| Wellcome Trust DBT India Alliance | 44 |
| DBT-ICGEB Centre for Advanced Bioenergy Research, New Delhi | 34 |
| Biotechnology Industry Research Assistance Council (BIRAC) | 17 |
| Bharat Immunological & Biological Corporation | 1 |
| **Indian Institute of Science Education and Research (IISER) System** | |
| IISER Pune | 3306 |
| IISER Kolkata | 3127 |
| IISER Bhopal | 1857 |
| IISER Mohali | 1836 |
| IISER Thiruvananthapuram | 1090 |
| IISER Tirupati | 340 |
| **Indian Council of Medical Research (ICMR) System** | |
| ICMR - National Institute of Cholera & Enteric Diseases (NICED) | 1222 |
| ICMR - National Institute of Nutrition (NIN) | 984 |
| ICMR - National Institute for Research in Tuberculosis (NIRT) | 935 |
| ICMR - National Institute of Malaria Research (NIMR) | 877 |
| ICMR - National Institute of Virology (NIV) | 727 |
| ICMR - National Institute for Research in Reproductive Health (NIRRH) | 704 |
| ICMR - Vector Control Research Center (VCRC) | 596 |
| ICMR - National Institute of Immunohaemotology (NIIH) | 542 |
| ICMR - National AIDS Research Institute (NARI) | 434 |
| ICMR - National Institute of Cancer Prevention & Research (NICPR) | 385 |
| ICMR - National Institute of Pathology (IOP) | 368 |
| ICMR - National JALMA Institute for Leprosy & Other Mycobacterial Diseases, Agra | 350 |
| ICMR - Rajendra Memorial Research Institute of Medical Sciences (RMRI) | 347 |
| ICMR - Regional Medical Research Centre (RMRC), Bhubaneswar | 338 |
| ICMR - National Institute of Epidemiology (NIE) | 289 |
| ICMR - National Institute of Occupational Health (NIOH) | 246 |
| National Centre for Disease Control (NCDC) | 232 |
| ICMR - National Institute for Research in Tribal Health (NIRTH) | 204 |
| ICMR - National Institute of Traditional Medicine (NITM) | 167 |
| ICMR - Regional Medical Research Centre, Port Blair (RMRCPB) | 164 |
| ICMR - Virus Unit, Kolkata | 145 |
| ICMR-VCRC Field Station, Madurai | 122 |
| ICMR - National Institute of Medical Statistics (NIMS) | 121 |

| Institution | Count |
|---|---|
| ICMR - Bhopal Memorial Hospital & Research Center (BMHRC) | 112 |
| ICMR - Regional Medical Research Centre, North East Region (RMRC NE) | 112 |
| ICMR - National Institute for Research in Environmental Health (NIREH) | 95 |
| ICMR - National Animal Resource Facility for Biomedical Research (NARFBR) | 72 |
| ICMR - National Institute for Implementation Research on Non-Communicable Diseases (NIIRNCD) | 60 |
| ICMR - National Centre for Disease Informatics & Research (NCDIR) | 51 |
| ICMR - Enterovirus Research Centre | 41 |
| ICMR - Regional Medical Research Center, Gorakhpur (RMRCGKP) | 19 |
| **Ministry of Earth Sciences (MoES) System** | |
| Indian Institute of Tropical Meteorology (IITM) | 1920 |
| India Meteorological Department (IMD) | 982 |
| National Centre for Polar and Ocean Research (NCPOR) | 650 |
| National Institute of Ocean Technology (NIOT) | 634 |
| Indian National Centre for Ocean Information Services (INCOIS) | 438 |
| National Centre for Medium Range Weather Forecasting (NCMRWF) | 352 |
| National Centre for Earth Science Studies (NCESS) | 328 |
| National Centre for Coastal Research (NCCR) | 244 |
| Centre for Marine Living Resources & Ecology (CMLRE) | 157 |
| National Center for Seismology (NCS) | 81 |
| **Indian Institute of Management (IIM) System** | |
| IIM Ahmedabad | 697 |
| IIM Bangalore | 502 |
| IIM Calcutta | 501 |
| IIM Lucknow | 298 |
| IIM Indore | 267 |
| IIM Kozhikode | 251 |
| IIM Raipur | 166 |
| IIM Udaipur IMU | 148 |
| IIM Rohtak | 82 |
| IIM Ranchi | 71 |
| IIM Kashipur | 66 |
| IIM Tiruchirappalli | 60 |
| IIM Shillong | 35 |
| IIM Amritsar | 27 |
| IIM Sambalpur | 20 |
| IIM Nagpur | 17 |
| IIM Sirmaur | 17 |
| IIM Visakhapatnam | 12 |
| IIM Jammu | 7 |
| IIM Bodh Gaya | 4 |
| **National Institute of Pharmaceutical Education & Research (NIPER) System** | |
| NIPER S.A.S. Nagar (Mohali) | 2260 |
| NIPER Hyderabad | 435 |
| NIPER Ahmedabad | 22 |
| NIPER Guwahati | 5 |
| NIPER Hajipur | 4 |
| NIPER Kolkata | 3 |
| NIPER Raebareli | 3 |